\documentclass[buckow,psfrag]{article}

\newcommand{\beqn}   {\begin{eqnarray}}
\newcommand{\eeqn}   {\nonumber \end{eqnarray}}

\newcommand{\eps}   {\epsilon}

\newcommand{\Dslash}{\mbox{$D$ \kern-.92em  \big /}}

\usepackage{amsmath,epsf,color,colordvi,shadow,pifont}
\usepackage{amsfonts}
\usepackage{epsfig,rotating}
\usepackage{amssymb}
\usepackage{psfrag}
\usepackage{buckow}

\begin {document}

\thispagestyle{empty}
\begin{flushright}
SPIN-01/27 \\
ITP-UU-01/36 \\
HU-EP-01/57 \\ 
hep-th/0112015 \\ 
\end{flushright}
\def\titleline{
Intersecting Brane Worlds \\[.4cm]
on Tori and Orbifolds 
}
\def\authors{
Ralph Blumenhagen\1ad, Boris K\"ors\2ad, Dieter L\"ust\1ad, 
and Tassilo Ott\1ad}
\def\addresses{
\1ad
Humboldt-Universit\"at zu Berlin, Institut f\"ur Physik \\ 
D-10115 Berlin, Germany\\
{\tt e-mail: blumenha, luest, ott@physik.hu-berlin.de} \\[.3cm]

\2ad 
Spinoza Institute, Utrecht University \\
Utrecht, The Netherlands \\
{\tt e-mail: kors@phys.uu.nl}
}
\def\abstracttext{
We review the construction and phenomenology of intersecting brane worlds. 
The breaking of chiral and supersymmetry together with a 
reduction of gauge symmetry by introducing D-branes at generic angles 
into type I or type II string compactifications is taken 
as a starting point for the engineering of phenomenologically 
appealing four-dimensional vacua. Examples that come very close to 
Standard Model and GUT physics are considered and issues like the 
cancellation of anomalies and the emergence of global symmetries studied 
in some detail. Concerning the gravitational backreaction, 
the perturbative potential generated at the 
disc level is shown to lead to a run-away instability of geometric 
moduli within the purely 
toroidal compactifications. It can at least partly be avoided in certain 
orbifold vacua, where the relevant moduli are frozen. These more 
restricted models still possess sufficient freedom to find   
semi-realistic brane world orbifolds.  
}
\large
\makefront

\section{Introduction}

The central issue in the construction of four-dimensional vacua 
with interesting low-energy dynamics from ten-dimensional string theory is 
the breaking or reduction of the various symmetries involved. 
Upon a Kaluza-Klein reduction on an internal toroidal space the 
ten-dimensional theories give rise to 16 or 32 supersymmetries,  
a gauge group of rank 16 and only nonchiral fermion spectra. 
As a novel approach to make contact with low energy phenomenology 
within type II or type I string theory, the concept of 
intersecting brane worlds has been developed. The gauge fields are confined 
to a three-dimensional 
subspace in the entirely six-dimensional internal space, the location 
of D6-branes, and in order to obtain four-dimensional vacua with chiral 
fermion spectra in semi-realistic gauge groups and with 
little or no supersymmetry these intersect each other in certain patterns. 
The idea dates back, in a dual disguise, to the work 
\cite{hepth9503030} which was the first to realize explicitly, that 
certain magnetic background fields can break supersymmetry, together with 
chiral and gauge symmetry. It was then embedded into string theory 
for toroidal and orbifold background spaces \cite{hepth0007024,hepth0007090} 
and generalized to include also the 
NSNS $B$ field modulus \cite{hepth0010279,hepth0012156} 
(see also \cite{K2001} for a review). 
The background fluxes, when interpreted as 
a constant background for open string propagation, induce a modification 
of the standard Dirichlet boundary conditions, which maps to the 
boundary conditions of rotated D-branes \cite{hepth9606139} under T-duality. 
This duality also implies a mapping between commutative and noncommutative 
background spaces for open strings ending on D$p$-branes with 
magnetic background fields and such ending on D$(p-1)$-branes 
located at relative angles as well as an identification of symmetric 
and asymmetric orbifolds \cite{hepth0003024}. \\   

In the resulting picture one deals with intersecting brane worlds on 
D6-branes, where 
the effective theory on the noncompact four-dimensional space common to all 
the D6-branes is determined by the intersection patterns on the 
internal torus or orbifold. 
A systematic analysis of the phenomenological 
properties and the perspectives to engineer a Standard Model in this approach 
was performed in \cite{hepth0007024,hepth0011073,hepth0012156,hepth0107138}. 
It is an interesting feature of these models that the
standard model Higgs field can be possibly identified with an open
string tachyonic field in a bifundamental representation.
The general philosophy of breaking supersymmetry\footnote{The 
techniques of intersecting brane worlds have also been 
employed to obtain vacua which still preserve ${\cal N}=1$ supersymmetry in 
four dimensions \cite{hepth0107166}. 
The resulting compactifications are believed 
to have purely geometrical interpretation in eleven-dimensional M-theory 
given by manifolds with holonomy $G_2$.} 
already at the 
string scale follows the so-called large extra dimensions scenario 
\cite{hepph9803315}.\\
 

In the absence of supersymmetry the cancellation of forces either 
among the branes themselves and also with respect to their effect onto 
the background geometry is no longer valid. As a first approximation 
to the true dynamics one can compute the leading 
perturbative contribution of the disc diagram 
to the potential of the background K\"ahler and complex structure moduli 
\cite{hepth0107138}. It turns out that in purely toroidal models the 
tension of the D6-branes always exceeds that of orientifold planes, if present 
at all, which implies the impossibility of partial 
supersymmetry breaking. The potential actually pushes the complex 
structure moduli to a degenerate limit where the torus is completely 
squashed. The generic strategy to deal with such situations would 
be to perform a shifting of the background in the way of a Fischler-Susskind 
mechanism \cite{FS1986} 
which is however very demanding and practically not 
tractable. A more modest method to evade the run-away behavior is given 
by freezing the relevant moduli through orbifolding. We follow this 
route by using a ${\mathbb{Z}}_3$ orbifold background together with 
the intersecting brane world approach, where the disc potential for all 
the moduli, K\"ahler, complex structure as well as twisted, is flat or 
fixed. The more severe restrictions which are imposed by the orbifolding 
still leave enough freedom to construct a class of models very close to the 
Standard Model, albeit with a number of problems in the Higgs sector. 

\section{Intersecting D-branes} 

In this preliminary section we collect the ingredients to the construction of 
intersecting brane worlds without going into greater detail of their 
derivation. The starting point is given by the observation that the 
string spectrum at any intersection point of two D-branes is given 
by a single fermion of some definite chirality, the ground state
in the R sector, together with scalars from the NS sector,  
which are generically massive, but may 
also be tachyonic, i.e. of negative mass squared. The tachyonic scalars 
may be 
related to Higgs fields which mediate spontaneous symmetry breaking 
in the effective theory whereas the massive states decouple from the massless 
sector. All of these fields come in bifundamental representations of the gauge 
group $U(N_a)\times U(N_b)$ where the $N_a$ and $N_b$ are the respective 
numbers of branes at the intersection. 
In addition, on any brane $a$ there is also the usual 
massless ${\cal N}=4$ vectormultiplet in the adjoint of the $U(N_a)$ 
from the strings with both ends on that brane. \\  

The classification of supersymmetric brane configurations can most 
easily been performed by evaluating the Killing spinor equations for 
any one brane and a second rotated one:  
\beqn
\Gamma_{0...6} \eps = \tilde\eps\quad \& \quad R \Gamma_{0...6} R^{-1} \eps=
\tilde\eps\ \Rightarrow R^2 = 1 
\eeqn
with $R = \exp \left( \varphi_{ab} \Gamma_{ab} /2  \right)$ the rotation 
operator and $\varphi_{ab}$ the relative angles of branes $a$ and $b$. 
Supersymmetry is preserved whenever $R \in SU(3)$. This 
is in a one to one correspondence with the BPS no-force law
\beqn 
{\cal A}_{ab} = 
\int_0^\infty{dl\ \langle B_a \vert e^{-l {\cal H}_{\rm cl}} \vert
  B_b \rangle} \sim ( 1_{\rm NSNS} - 1_{\rm RR} ) , 
\eeqn 
here written for the balance of attractive and repulsive forces in the annulus 
amplitude. Therefore, only supersymmetric configurations can lead to globally 
stable brane worlds, at least if no other background fields are 
being switched on. 
As mentioned already, in order to avoid the common problem of stabilizing 
the hierarchy of gravitational and gauge theoretical scales in the absence 
of supersymmetry, one needs to take reference to a large extra dimension 
scenario, where the internal space transverse to the D-branes is large 
compared to the space along the branes. 
As has been noticed in \cite{hepth0007024} in the purely 
toroidal approach, there is no overall internal transverse space that could 
be chosen large in order to realize such a situation, such that one needs 
to depart from the flat background to do so. We shall later mention 
the opportunity to blow up orbifold fixed points as one such possibility. 
In any case one still needs to consider the stability of the geometrical 
hierarchy of volumes, once fixed at tree level, when perturbations are taken 
into account, as well as the appearance of 
open string tachyons.\footnote{See also 
\cite{hepth0107036} in this context.} \\ 

The above completely generic features of the construction get slightly 
modified and extended when turning to type I (or type I$'$ 
after the T-duality) string theory.  
One needs to add orientifold 
O6-planes and gets additional contributions to the 
Euler characteristic zero amplitude via the Klein bottle and M\"obius strip 
diagrams. 
As well, one needs to take into account that the world sheet parity 
reflects the magnetic background fields of type I, which translates into 
a geometric reflection symmetry $\Omega{\cal R}$ acting as $\varphi_{a} 
\longleftrightarrow -\varphi_{a}$ on the relative angles of the 
D-brane spectrum. 

\section{Intersecting brane worlds} 

Upon compactifying a setting of intersecting D6-branes on a six-dimensional 
internal torus ${\mathbb{T}}^6 = \bigotimes_{I=1}^3{\mathbb{T}}_I^2$, 
factorized 
into two-dimensional ones with K\"ahler moduli $T^I$ and complex 
structure moduli $U^I$, any such D6-brane wraps a special Lagrangian 
3-cycle given by a line on any single ${\mathbb{T}}^2_I$. The homology 
class of the 3-cycle is then defined 
by combining those of the 1-cycles, i.e. by 
three sets of two integers, one set 
for each ${\mathbb{T}}^2_I$, as being depicted for two examples 
of tori with different complex structures in the figure below. 
\begin{center}
\psfrag{nm21}[bc][bc][.7][0]{$(n_b,m_b)=(1,-1)$} 
\psfrag{nm22}[bc][bc][.7][0]{$(n_c,m_c)=(2,-1)$} 
\psfrag{nm12}[bc][bc][.7][0]{$(n_a,m_a)=(1,1)$} 
\psfrag{i}[bc][bc][.7][0]{$i$} 
\psfrag{1}[bc][bc][.7][0]{1} 
\psfrag{U}[bc][bc][.7][0]{$-1/U$} 
\psfrag{B}[bc][bc][.7][0]{$B'=0$} 
\psfrag{Bn}[bc][bc][.7][0]{$B'\not=0$} 
 \epsfxsize=8cm
 \epsfysize=4cm
 \epsfbox{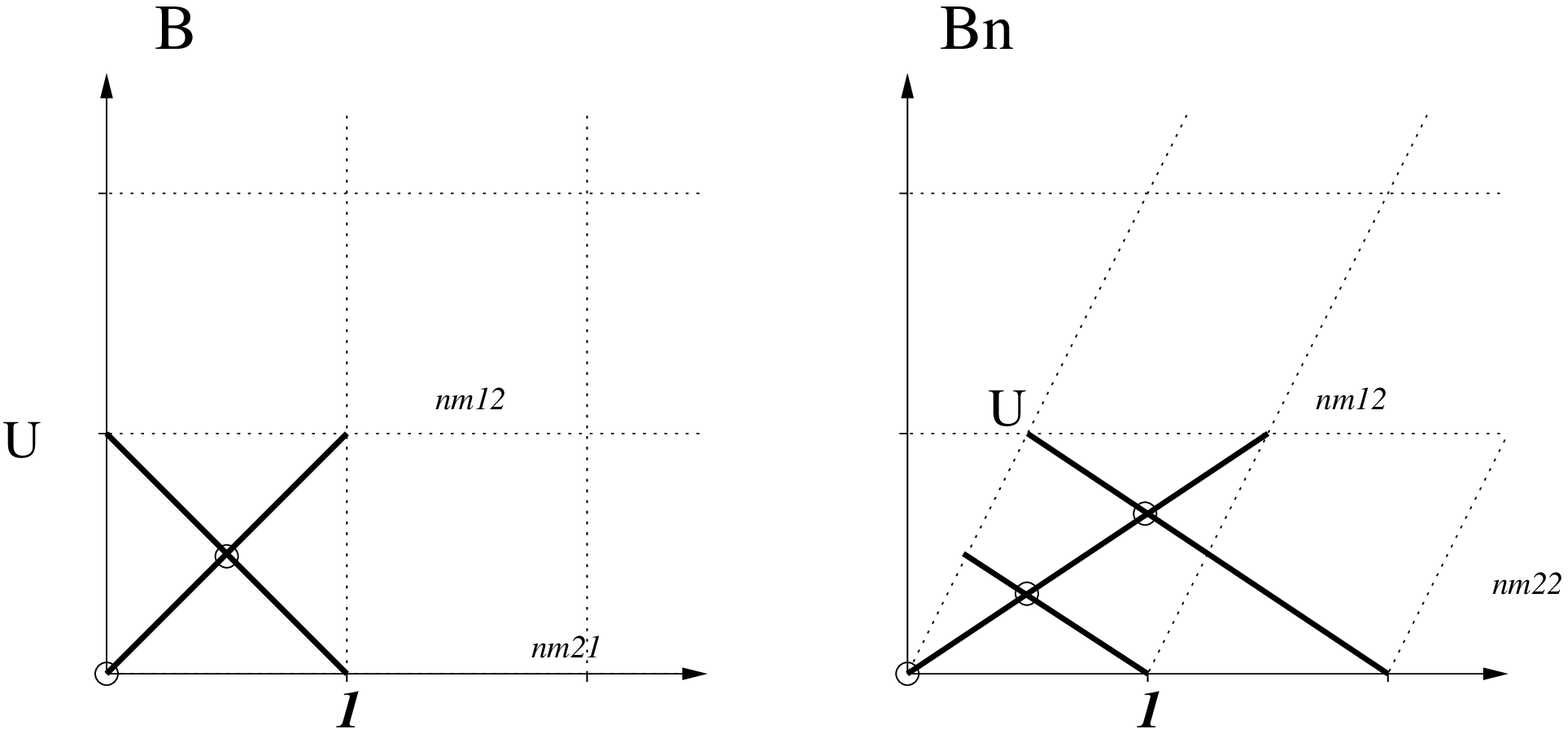}
%
\end{center}
The number of chiral fermions $\chi$ 
in the $({\bf N}_a,{\bf N}_b)$ representation 
is given by the intersection number 
\beqn
\chi\ {\rm in}\ ({\bf N}_a,{\bf N}_b):\quad I_{ab} = \prod_a{\left( m^I_a
    n^I_b-m^I_a n^I_b \right)} 
\eeqn
of the two respective cycles. A crucial step in the development of 
semi-realistic type I models of this kind was to realize that the intersection 
numbers depend on the value of $\Re(U^I)=0,1/2$, odd numbers being 
accessible only with non-vanishing $\Re(U^I)$ \cite{hepth0012156}.
Along with the appearance of chiral fermions one 
needs to address the consistency requirements of anomaly cancellation 
in the effective theory, which lifts to the cancellation of RR charge 
in string theory. An explicit computation of the conditions that 
result in the different theories can be found in the references. 
The conditions which emerge for instance for type I strings on a 
purely toroidal background are given by \cite{hepth0007024}  
\beqn
\sum_a{N_a n^1_a n^2_a n^3_a} = 32 ,\ \sum_a{N_a n^1_a m^2_a m^3_a} = 
\sum_a{N_a m^1_a n^2_a m^3_a} = \sum_a{N_a m^1_a m^2_a n^3_a} =0 . 
\eeqn
They have been shown to imply the cancellation of all irreducible 
contributions to the anomalies in four dimensions. \\ 

As an example for a solution with interesting features, 
we present a unified extension of the 
Standard Model, a left-right symmetric Standard Model in a type I 
intersecting brane world. 
The brane setting can be read off from the 
following figure 
\begin{center}
\hspace{0cm}
{
\psfrag{1}[bc][bc][.7][0]{D$_1$}
\psfrag{2}[bc][bc][.7][0]{D$_2$}
\psfrag{3}[bc][bc][.7][0]{D$_3$}
\psfrag{4}[bc][bc][.7][0]{D$_4$}
 \epsfxsize=9cm
 \epsfysize=3cm
 \epsfbox{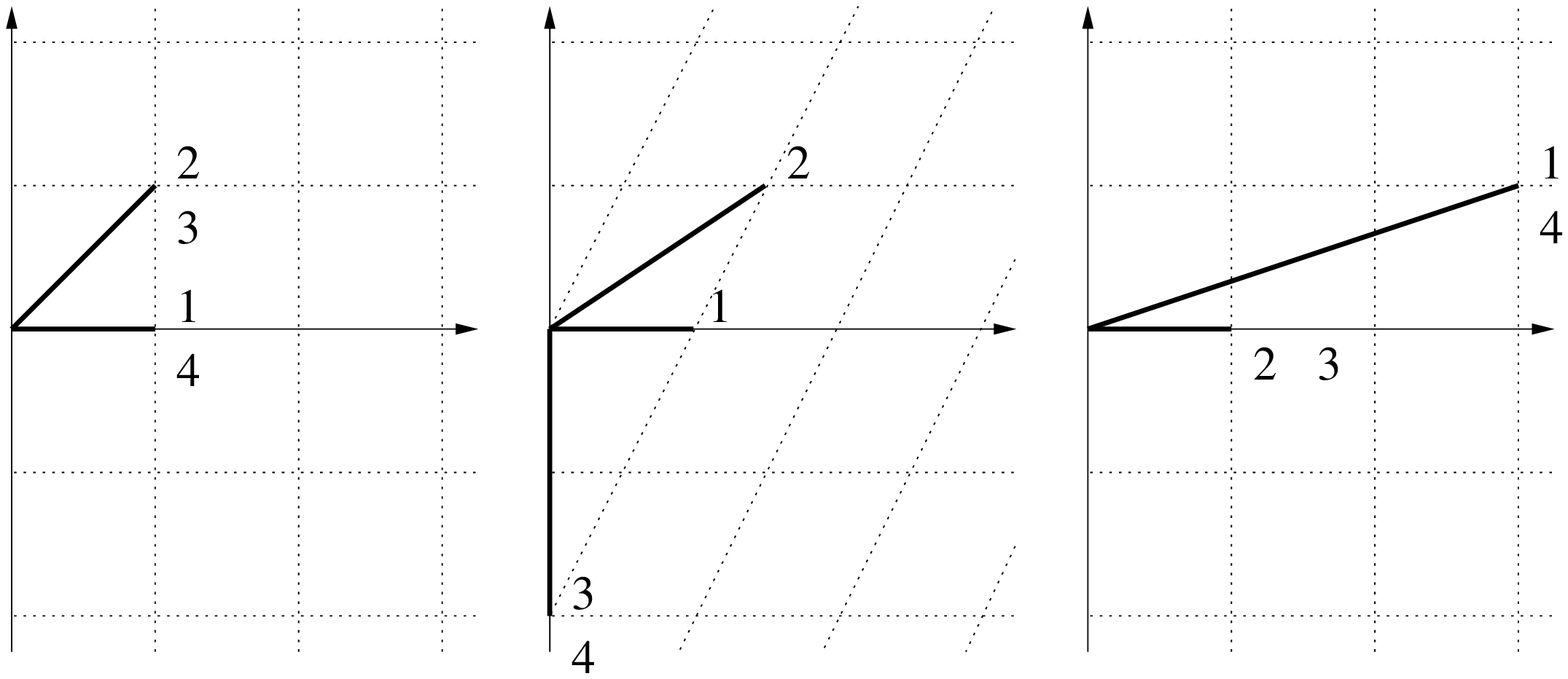}
}
%
\end{center}
The RR charges cancel for the choice 
$N_1 =3,\ N_2=N_3=2,\ N_4=1$ 
for the respective multiplicities of branes, leading to the gauge group  
\beqn 
SU(3)_c\times SU(2)_L\times SU(2)_R \times U(1)_{B-L} \times U(1) . 
\eeqn
One needs to take care that of the originally 
four $U(1)$ factors only two are free
of anomalies, whereas the other two decouple at the string scale, due 
to Green-Schwarz type interactions. 
The spectrum of massless fermions is tuned to produce the desired 
set of three generations of quark and lepton doublets, as is usually 
considered within GUT models of this kind. \\ 

Many of the phenomenological properties of the effective theory can be 
discussed without explicit reference to a particular model. 
The gauge coupling constants can be obtained from ten dimensions 
by a straightforward dimensional reduction along the world volume of 
the respective brane from the string coupling, i.e. 
\beqn
\frac{1}{g^{2}_a} \sim \frac{1}{g_{\rm S}} \frac{{\rm vol}_a}{l^3_{\rm S}} , 
\eeqn 
where ${\rm vol}_a$ is the volume of the brane $a$. 
These extra relative factors may change the unification patterns 
compared to standard field theory results.  
The Yukawa couplings involve fields that live at different intersection 
loci on the internal space and therefore arise from string diagrams with 
world sheets that 
stretch between different branes. 
%
The coupling strength can then be estimated classically by the minimal area, 
e.g. $\exp(-{\rm A}_{\bar\psi \psi H}/l_{\rm S}^2)$. 
These interactions are suitable to generate mass terms for the nonchiral part 
of the matter spectrum, such as the fermions in the ${\cal N}=4$ 
vectormultiplets, which gain masses 
\beqn 
M_{\bar\psi \psi}^2 \sim g_{\rm Yu} \langle H \rangle \sim l_{\rm S}^{-2}
\eeqn
of the order of the string scale and decouple. 
It was argued in \cite{hepth0011073} that there may be a specific signature 
in this class of large volume compactifications that may be able 
distinguish them from other kinds of models, excitations of 
fields localized at the intersections: 
\beqn
\left( \alpha_{-\delta_{ab}} \right) ^n \vert 0 \rangle \quad {\rm with}\quad 
M_n^2 \sim n q \langle \delta_{ab} \rangle = \frac{nq}{\pi} \langle 
\varphi_{ab} \rangle \stackrel{?}{\ll} l_{\rm S}^{-2} . 
\eeqn 
The appearance of anomalous $U(1)$ symmetries, cured by a suitable version 
of the Green-Schwarz mechanism, presents a convenient source for global 
symmetries.  
For instance, in type I string theory, there are four axionic 
scalars that derive from the RR forms $C^{(p)}$:   
\beqn 
4d\ {\rm axions}\quad a^I = \int_{\mathbb{T}^2_I}{C^{(2)}},\ a^0 =
\int_{\mathbb{T}^6}{C^{(6)}} . 
\eeqn 
The four-dimensional couplings relevant for the GS mechanism descend from  
$C^{(2)} \wedge {\rm tr}\, F^4$ and $C^{(6)} \wedge {\rm tr}\, F^2$, 
leading to an anomalous contribution schematically given by
\beqn 
N_a I_{ab}\sum_\alpha{\left( F \wedge *a^\alpha \right) \times \left( a^\alpha
    F\wedge F\right) } . 
\eeqn
The coefficient is just correct to 
match up with the contributions of the 
chiral fermions in order to make the GS mechanism work out. 
Those gauge bosons which belong to anomalous 
$U(1)$ gauge symmetries acquire mass terms of the order of the string scale, 
i.e. up to four abelian factors decouple, but they survive as global 
symmetries of the theory. Actually, one needs to be even more careful here 
and distinguish the abelian and mixed anomalies and also regard that 
non-anomalous gauge fields may get massive. In any case, this set of 
global symmetries may rather generically contain baryon- and/or 
lepton-number conservation, which prevents proton decay, otherwise mediated 
via dimension six four-fermion terms 
$l_{\rm S}^2 \left( \overline{u}^c_L u_L \right) ( \overline{e}^+_L d_L )$ 
in the effective action. In scenarios 
with low string scale, these would not be sufficiently suppressed, 
which usually is a major problem. 

\section{Gravitational stability and stabilization} 

So far, we have ignored the problem of the stability of our brane worlds.  
Now we proceed to consider the backreaction of the torus when the branes 
on top are arranged in a nonsupersymmetric way \cite{hepth0107138}. 
From the annulus diagram one 
can obtain the disc tadpoles $\langle \phi \rangle_{\rm disc}$ and 
$\langle U^I \rangle_{\rm disc}$ of the dilaton and complex 
structure moduli fields by 
taking the limit of an infinitely long and thin tube. 
This allows to deduce the scalar potential to this order of 
perturbation theory: 
\beqn 
V(\phi,U^I) = e^{-\phi} \left( \sum_a{N_a \prod_I{\sqrt{ (n_a^I)^2
   \Im(U^I) +
        \frac{(m^I_a + \Re(U^I) n_a^I)^2}{\Im(U^I)}}} }
- 16 \prod_I \sqrt{\Im(U^I)} \right) .  
\eeqn   
It is just the difference of the internal volumes of all the D6-branes and 
that of the orientifold O6-planes, which is also what one would have expected 
as the result of the Dirac-Born-Infeld effective action, evaluated for 
the present constant background. There is no potential for the K\"ahler 
moduli at this order, which is consistent with the fact that they do
 not couple to the 
boundary state that describes the D6-branes in the world sheet CFT. 
First of all, one can easily realize that there are no supersymmetric 
vacua where the potential vanishes together with its derivative. 
The volume of the D-branes is 
always larger than that of the orientifold planes. 
In fact, the potential does not allow for any nontrivial minima at all, 
but pushes the complex structure moduli $\Im(U^I)$ to infinity, while 
$\Re(U^I)$ is frozen anyway. Intuitively speaking, the D-branes are dragged 
towards the orientifold planes, 
and pull the entire 
torus to the degenerate limit, where the ratio of the radii diverges at 
constant volume. 
In the original T-dual type I picture this is the decompactification limit 
of the internal space. 
\begin{center}
\hspace{0cm}
{
\psfrag{U}[bc][bc][.7][0]{$-1/U$}
 \epsfxsize=12cm
 \epsfysize=3cm
 \epsfbox{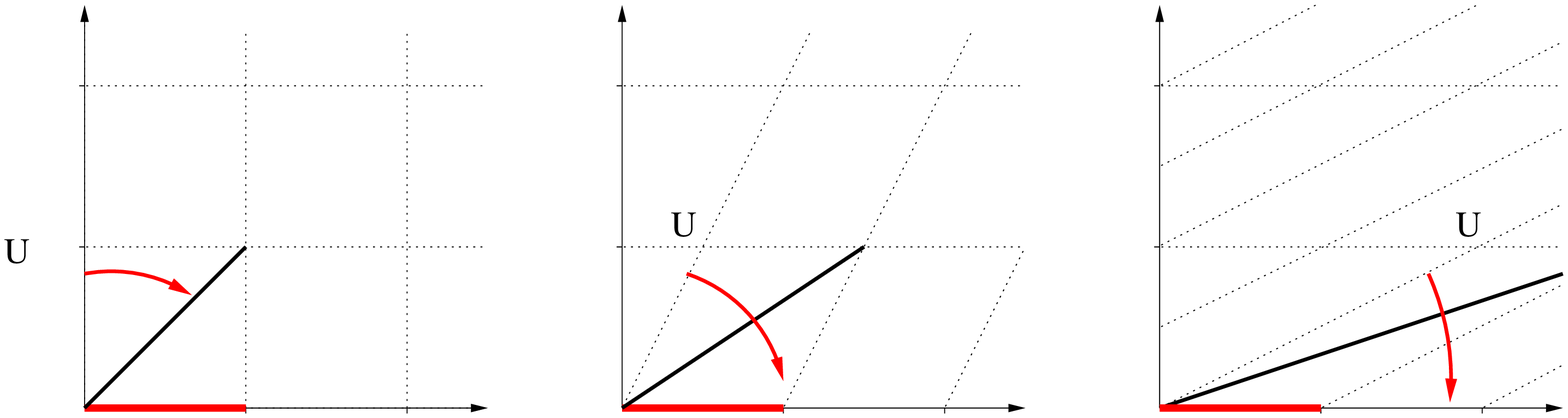}
}
%
\end{center}
The most generic strategy to deal with the tadpoles for the dilaton 
and the moduli would be to follow the Fischler-Susskind mechanism 
\cite{FS1986}, 
which is however not practicable and very few examples are known where 
even only the first step can be performed \cite{hepth0004165}. 
A more modest approach consists in freezing the relevant moduli by 
imposing a certain discrete 
symmetry on the background via orbifolding. More precisely, 
a ${\mathbb{Z}}_3$ orbifold of type I theory allows only the two 
distinct values 
$U^I =(1+i\sqrt{3})/2$ or $(1+i/\sqrt{3})/2$ 
for the complex structure modulus. The orientifold compactification 
relevant here is based on a type IIA orbifold and has been constructed in 
\cite{hepth9912204}. Interestingly, there are no twisted 
tadpoles generated and therefore  the potential is entirely flat except for 
the dilaton that runs away to zero coupling:
\beqn
V(\phi) = e^{-\phi} \cdot {\rm const.} 
\eeqn
The fact that the branes do not couple to the twisted moduli that 
parametrize the blowing-up of the singularities suggests a nice 
realization of a large extra dimension scenario, where the internal 
volume transverse to the branes appears when such a singularity is blown 
up. Of course, one has to expect a potential for these moduli from 
higher loop corrections and, thus, this argument remains rather heuristic.  
In an explicit construction of such models\footnote{For 
proper definitions and a presentation of 
the technical details of this subsection 
see \cite{hepth0107138}.} 
one now needs to employ D-branes 
intersecting in patterns invariant under the additional 
${\mathbb{Z}}_3$ rotation.  
Thus, any single D-brane is a representative of an entire orbit 
of equivalent 
branes, generically of length six, sometimes only of length three.   
One can very efficiently redo the former toroidal analysis of the 
chiral spectrum and the RR charge cancellation constraints by introducing 
effective winding numbers $(z_a,y_a)$ 
which summarize the contributions of the individual branes into brane orbits. 
The only condition one finds is 
\beqn
\sum_a{N_a z_a} - 2 =0 . 
\eeqn
It actually does not prevent a rank of the gauge group larger than two, as 
opposed to the supersymmetric solutions of \cite{hepth9912204}. 
A promising example for a model of this type that resembles the Standard Model 
with respect to gauge group and spectrum very closely is defined by 
three stacks of D6-branes with 
$(z_1,y_1) = (z_2,y_2) = (1/2,3), \ (z_3,y_3) = (-1/2,3)$. 
It has gauge group 
\beqn
SU(3)_c \times SU(2)_L \times U(1)_Y \times U(1)_{B-L} 
\eeqn
after applying the GS mechanism and a fermion spectrum identical 
to the Standard Model with right-handed neutrinos. 
The Higgs fields, necessary to break $U(1)_{B-L}$ as well as
$SU(2)_L\times U(1)_Y$,
should correspond to open string tachyons with the correct gauge 
quantum numbers.
By a systematic computer search among all the $4\cdot 36^3$ solutions with 
these characteristics, which we were able to find, there 
were a couple of hundred models having either a standard
model tachyonic Higgs in
the $({\bf 1},{\bf 2})$ representation or a Higgs in the `symmetric'
representation of $U(1)_{B-L}$. But no model contains both Higgs fields.
As an additional problem the global symmetry, 
which orginates from the single anomalous and 
decoupled $U(1)$ gauge factor prevents the 
coupling $H\, \overline{Q}_L u_R$ needed to give mass to some of the 
quarks during the electroweak symmetry breaking phase transition. Therefore, 
we can finally only refer to some version of an exotic Higgs mechanism 
possibly involving a composite field that takes over the role 
of the missing elementary scalar. \\ 

In any case, the intersecting brane world 
approach offers a rich phenomenology with convenient tools for 
engineering vacua with many attractive features. We have also 
shown here 
that it is possible to address the very serious issue of stabilizing 
nonsupersymmetric brane world vacua by a perturbative CFT approach based 
on orbifold techniques. This work is surely not complete and many more 
perspectives remain for the future.


\end{document}